\documentclass[prb,preprint,amsmath,amssymb,superscriptaddress]{revtex4}
\usepackage{graphicx}
\usepackage{dcolumn}
\usepackage{bm}
\usepackage{natbib}
\usepackage{amssymb}
\usepackage{placeins}
\usepackage{epsfig}
\usepackage{xspace}
\usepackage{miller}
\newcommand{\etal}{\emph{et al}\xspace}
\newcommand{\cumnas}{CuMnAs\xspace}

\newcommand{\beq}[1]{
\begin{equation}
\label{e#1} }
\newcommand{\eeq}{
\end{equation}
}

\begin{document}

\title{Tetragonal phase of epitaxial room-temperature antiferromagnet CuMnAs}
\author{P. Wadley}
\email{peter.wadley@nottingham.ac.uk}
\thanks{Corresponding author}
\affiliation{Institute of Physics ASCR, v.v.i., Cukrovarnick\'a 10, 162 53 Praha 6, Czech Republic}
\affiliation{School of Physics and Astronomy, University of Nottingham, Nottingham NG7 2RD, United Kingdom}
\author{V. Nov\'ak}
\affiliation{Institute of Physics ASCR, v.v.i., Cukrovarnick\'a 10, 162 53 Praha 6, Czech Republic}
\author{R.~P.~Campion}
\affiliation{School of Physics and Astronomy, University of Nottingham, Nottingham NG7 2RD, United Kingdom}
\author{C. Rinaldi}
\affiliation{LNESS-Dipartimento di Fisica del Politecnico di Milano Via Anzani 42, 22100 Como, Italy}
\affiliation{Institute of Physics ASCR, v.v.i., Cukrovarnick\'a 10, 162 53 Praha 6, Czech Republic}
\author{X.~Mart\'{i}}
\affiliation{Institute of Physics ASCR, v.v.i., Cukrovarnick\'a 10, 162 53 Praha 6, Czech Republic}
\affiliation{Faculty of Mathematics and Physics, Charles University in Prague, Ke Karlovu 3, 121 16 Prague 2, Czech Republic}
\affiliation{Department of Materials Science and Engineering, University of California, Berkeley, California 94720, USA}
\author{H.~Reichlov\'a}
\affiliation{Institute of Physics ASCR, v.v.i., Cukrovarnick\'a 10, 162 53 Praha 6, Czech Republic}
\affiliation{Faculty of Mathematics and Physics, Charles University in Prague, Ke Karlovu 3, 121 16 Prague 2, Czech Republic}
\author{J. \v{Z}elezn\'y}
\affiliation{Institute of Physics ASCR, v.v.i., Cukrovarnick\'a 10, 162 53 Praha 6, Czech Republic}
\author{J.~Gazquez}
\affiliation{Institut de CiÃšncia de Materials de Barcelona, ICMAB-CSIC, Bellaterra, Spain} 
\author{M.~A.~Roldan}
\author{M.~Varela}
\affiliation{Departamento de F’sica Aplicada III, Universidad Compluense de
Madrid, Madrid, Spain}
\affiliation{Materials Science \& Technology Division, Oak Ridge National Laboratory, USA}
\author{D.~Khalyavin}
\author{S.~Langridge}
\affiliation{ISIS, Rutherford Appleton Laboratory, Harwell Science and Innovation Campus,Science and Technology Facilities  Council,  Oxon.  OX11 0QX, United Kingdom}
\author{D. Kriegner}
\affiliation{Institute of Semiconductor and Solid State Physics, University Linz, Altenbergerstr. 69, A-4040 Linz, Austria}
\author{F.~M\'aca}
\author{J.~Ma\v{s}ek}
\affiliation{Institute of Physics ASCR, v.v.i., Na Slovance 2, 182 21 Praha 8, Czech Republic}
\author{R. Bertacco}
\affiliation{LNESS-Dipartimento di Fisica del Politecnico di Milano Via Anzani 42, 22100 Como, Italy}
\author{V.~Hol\'{y}}
\affiliation{Faculty of Mathematics and Physics, Charles University in Prague, Ke Karlovu 3, 121 16 Prague 2, Czech Republic}
\author{A.~W.~Rushforth}
\author{K.~W.~Edmonds}
\author{B.~L.~Gallagher}
\author{C.~T.~Foxon}
\affiliation{School of Physics and Astronomy, University of Nottingham, Nottingham NG7 2RD, United Kingdom}
\author{J.~Wunderlich}
\affiliation{Institute of Physics ASCR, v.v.i., Cukrovarnick\'a 10, 162 53 Praha 6, Czech Republic}
\affiliation{Hitachi Cambridge Laboratory, Cambridge CB3 0HE, United Kingdom}
\author{T.~Jungwirth}
\affiliation{Institute of Physics ASCR, v.v.i., Cukrovarnick\'a 10, 162 53 Praha 6, Czech Republic}
\affiliation{School of Physics and Astronomy, University of Nottingham, Nottingham NG7 2RD, United Kingdom}

\begin{abstract}
   Recent studies have demonstrated the potential of antiferromagnets as the active component in spintronics devices. This is in contrast to their current passive role as pinning layers in hard disk read heads and magnetic memories. Here we report the epitaxial growth of a new high temperature antiferromagnetic material, tetragonal CuMnAs, which exhibits excellent crystal quality, chemical order and compatibility with existing semiconductor technologies. We demonstrate its growth on the III-V semiconductors GaAs and GaP, and show that the structure is also lattice matched to Si. Neutron diffraction shows collinear antiferromagnetic order with a high N\'eel temperature. Combined with our demonstration of room temperature exchange coupling in a CuMnAs/Fe bilayer, we conclude that tetragonal CuMnAs films are suitable candidate materials for antiferromagnetic spintronics. 
\end{abstract}

\maketitle

Large and bistable magnetoresistance signals have been observed in tunneling devices with an antiferromagnet (AFM) IrMn on one side and a non-magnetic metal on the other side of the tunnel barrier.\cite{Park:2010_a,Marti:2012_a} The work has experimentally demonstrated the feasibility of a spintronic concept\cite{Shick:2010_a,Jungwirth:2010_a,Cava:2011_a} in which the electronic device characteristics are governed by the staggered magnetisation axis in an AFM.  It has been shown that the AFM moments can be manipulated via an exchange coupled ferromagnet (FM)\cite{Park:2010_a,Marti:2012_a,Wang:2012_a} and that the AFM magnetoresistance signals can persist to room temperature.\cite{Wang:2012_a} No stray fields and the relative insensitivity to external magnetic fields are among the features which make AFMs attractive complements to the conventionally utilized FMs in the design of spintronic devices, as highlighted in a recent study of AFM linear chains of a few Fe atoms, \cite{Loth:2012_a} and in the 
demonstration of the concept of an AFM memory. \cite{Petti:2013_a} The possibility to design spintronic elements based on AFMs becomes even more attractive in the context of magnetic counterparts of conventional compound semiconductors, which may enable new devices combining spintronic and nanoelectronic functionalities.  In this paper we report the discovery of a new member of this family: tetragonal epitaxial CuMnAs, a room-temperature AFM whose in-plane lattice constant matches GaP or Si. 

Over the past two decades, the introduction of magnetism into common semiconductor hosts has driven  a number of new research areas in spintronics. Inspired by spintronics research and applications based on transition metal ferromagnets (FMs), the focus has been on magnetic counterparts of semiconductors with the FM order. (Ga,Mn)As and related (III,Mn)V compounds\cite{Ohno:1998_a,Jungwirth:2006_a,Dietl:2008_c} have become archetypes among these materials, resulting in discoveries of new spin-related physical phenomena and device functionalities. For example, the understanding of spin-orbit coupling phenomena has advanced due to experiments in (Ga,Mn)As, from the ohmic transport regime to new realizations in tunneling and Coulomb blockade devices.\cite{Gould:2004_a,Wunderlich:2006_a}  Low Curie temperatures have prevented the direct integration of (III,Mn)V FM semiconductors into spintronics technologies. Nevertheless, new phenomena discovered in (Ga,Mn)As have been subsequently reported in the room-
temperature metal FMs, relevant not only for basic science but also for practical spintronic applications.\cite{Moser:2006_a,Gao:2007_a,Park:2008_a}

\begin{figure}[ht!]
\includegraphics[width=0.45\columnwidth]{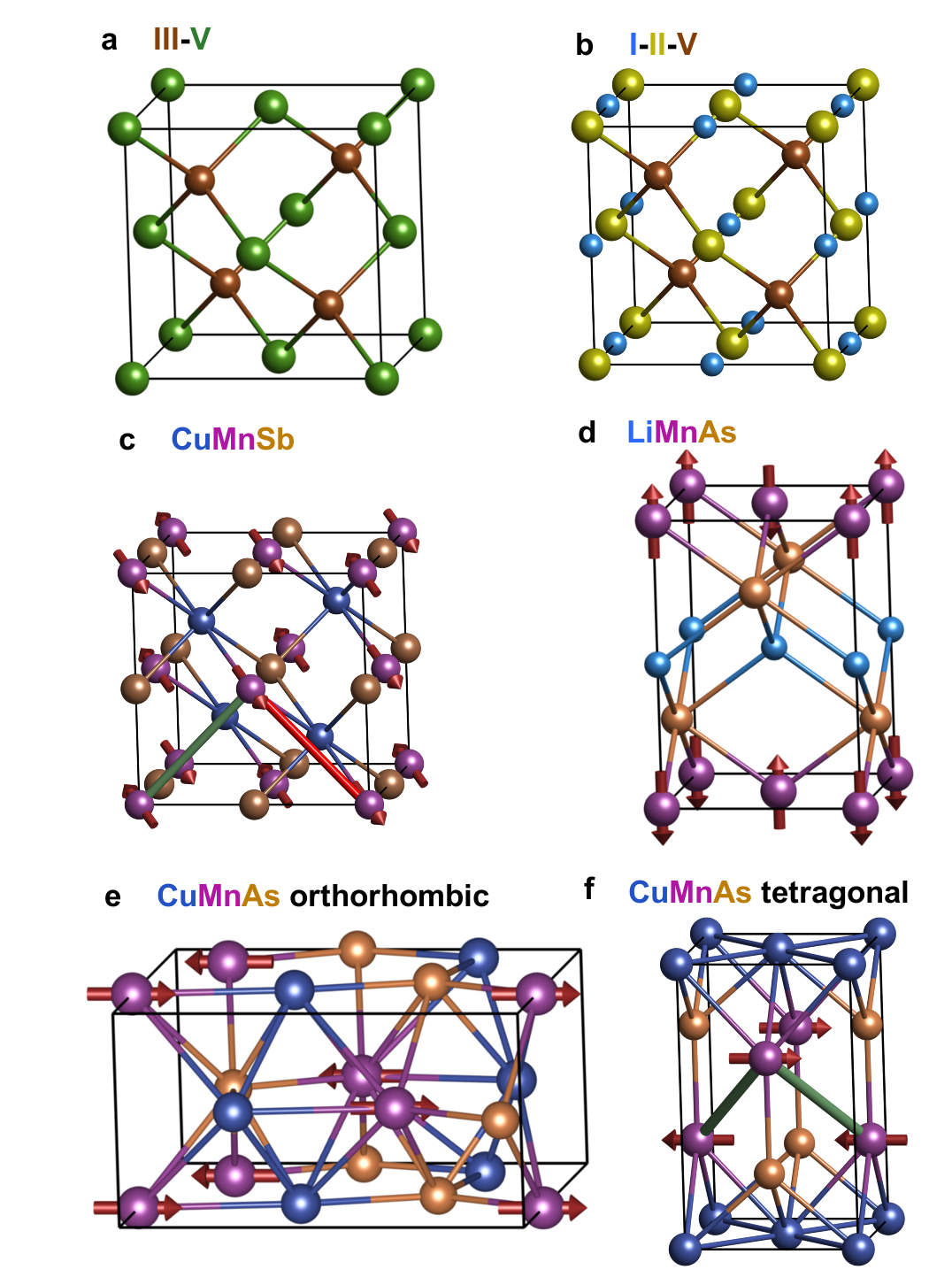}
\caption{\textbf{Unit cell structures of non-magnetic semiconductors and their magnetic counterparts.} \textbf{(a)} III-V zincblende structure. \textbf{(b)} I-II-V half-Heusler structure. \textbf{(c)} Half-Heusler CuMnSb. \textbf{(d)} Tetragonal LiMnAs. \textbf{(e)} Orthorhombic CuMnAs. \textbf{(f)} Tetragonal CuMnAs. The bonds in \textbf{c}, highlighted in green and red, show the AFM and FM nearest neighbor exchange coupling of Mn  present in the half-Heusler CuMnSb. The bonds in \textbf{f}, highlighted in green, show that all Mn nearest neighbors are coupled antiferromagnetically in the tetragonal CuMnAs which is favorable for high $T_N$.}
\label{fig:cryst}
\end{figure}

While a ferromagnetic ground state is rare within magnetic materials derived from semiconductor compounds and their Curie temperatures are below room temperature, high N\'eel temperature ($T_N$)  AFMs can be found, for example, among the magnetic counterparts of the I-II-V semiconductors (see also Supplementary Table S1 and Supplementary Note 1).\cite{Jungwirth:2010_a,Cava:2011_a}  The structure of the non-magnetic semiconductors such as LiZnAs, NaZnAs, CuZnAs and AgZnAs is closely related to the III-V zincblende structure, as shown in Fig.~\ref{fig:cryst}a,b. By splitting the group III atom into two elements from group I and II, their combined valence equals that of the group III atom. One of the two elements resides on the zincblende site while the other occupies one of the tetrahedral interstitial sites that is empty in the zincblende structure.\cite{Zunger:1993_a} An extensively studied magnetic compound with the same filled zincblende (also called half-Heusler) crystal structure is CuMnSb (Fig.~\ref{fig:cryst}c).
\cite{Forster:1968_a,Endo:1970_a} The AFM coupling of FM (111) planes of this cubic crystal has a frustration in the magnetic interactions, with half of the Mn nearest-neigbours coupled antiferromagnetically and half ferromagnetically.\cite{Maca:2012_a} This frustration leads to the relatively low $T_N=50$~K of CuMnSb and shows that keeping the same cubic structure as in the non-magnetic parent I-II-V compound is unfavourable for robust antiferromagnetism. Moreover, lowering the symmetry from cubic to, e.g., tetragonal enhances the magnetocrystalline anisotropy phenomena and is therefore favourable for the concept of AFM spintronics.

LiMnAs and NaMnAs are among the examples of AFM I-II-V semiconductors\cite{Bronger:1986_a,Muller:1991_a,Schucht:1999_a,Jungwirth:2010_a,Wijnheijmer:2012_a} whose equilibrium crystal structure changes from the cubic half-Heusler lattice of their non-magnetic counterparts to a layered tetragonal structure (Fig.~\ref{fig:cryst}d). This more anisotropic crystal arrangement removes the frustration in the nearest-neighbor magnetic coupling, resulting in $T_N$'s far above room temperature. Single crystal thin films of the LiMnAs AFM semiconductor were recently prepared by molecular beam epitaxy (MBE) on lattice matched InAs substrate.\cite{Jungwirth:2010_a,Novak:2011_a,Wijnheijmer:2012_a} The inclusion of alkali metal elements represents, however, a challenge both in terms of the growth and the stability of devices. Group Ib transition metal elements may represent the solution to this problem: NaZnAs and AgZnAs are known to be twin compounds with identical crystal structure and lattice constants, and CuZnAs is also 
very similar.\cite{Zunger:1993_a} This has motivated our interest in CuMnAs as a suitable I-Mn-V AFM compound for spintronics.\cite{Maca:2012_a} The bulk equilibrium phase of CuMnAs displays room temperature AFM ordering,\cite{Mundelein:1992_a} as confirmed in our recent study of chemically synthesized bulk samples.\cite{Maca:2012_a} However, the orthorhombic crystal structure (Fig.~\ref{fig:cryst}e) is not compatible with conventional compound semiconductor substrates. 

In the present work we demonstrate that CuMnAs films can be prepared by MBE in the tetragonal form reminiscent of NaMnAs (Fig.~\ref{fig:cryst}f). In analogy to the Ia-Mn-V compounds, all nearest-neighbours in the tetragonal CuMnAs films couple antiferromagnetically, resulting in a high $T_N$. We demonstrate relaxed epilayers of the tetragonal phase of CuMnAs grown on GaAs substrate and fully strained CuMnAs films grown on the closely lattice matched GaP. The functionality of this new AFM I-II-V compound is demonstrated in all-epitaxial CuMnAs/
Fe AFM/FM bilayers where we observe large room-temperature exchange bias effects.

\section*{Results}
\begin{figure}[ht!]
\includegraphics[width=0.90\columnwidth]{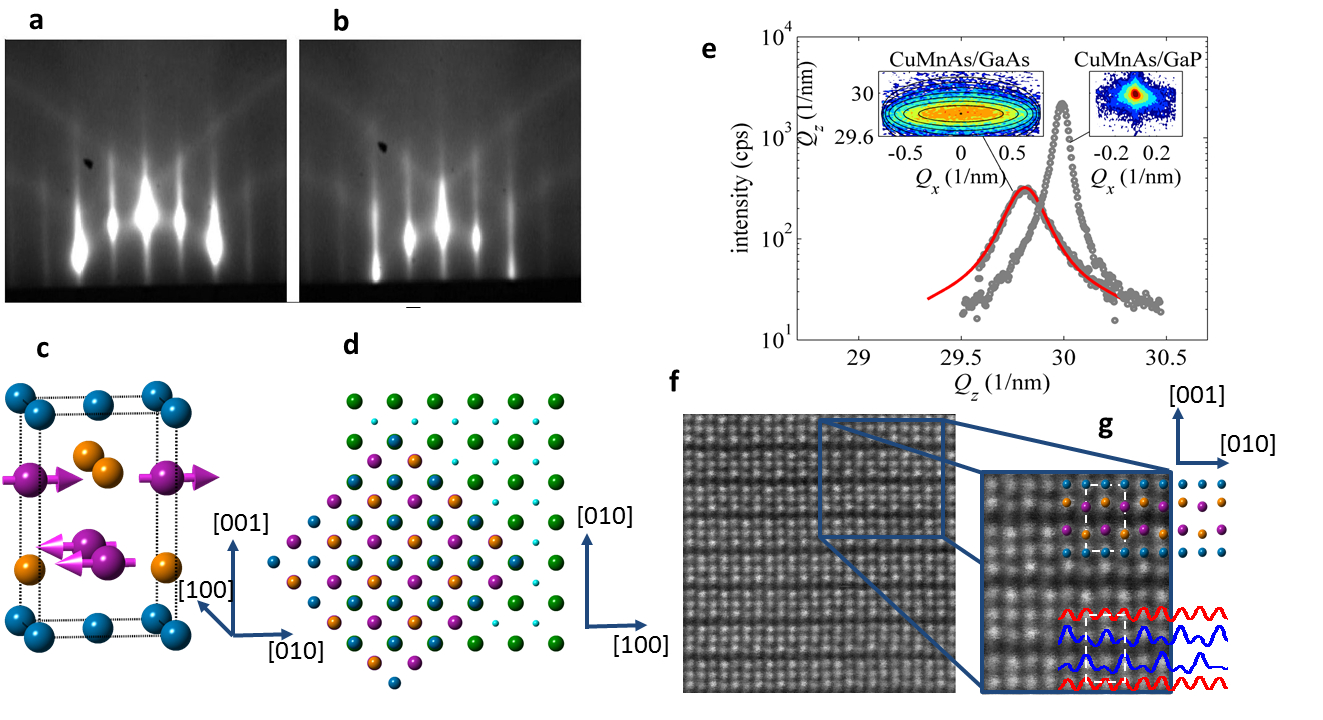}
\caption{\textbf{Growth and structural characterisation of CuMnAs.} \textbf{(a)} and \textbf{(b)} RHEED images of the surface of a ${\rm CuMnAs}$ layer during growth on a GaP substrate taken along the substrate \hkl[110] and \hkl[-110] respectively. \textbf{(c)} Diagram depicting the CuMnAs unit cell structure and spin arrangement, with Mn (purple), Cu(blue) and As(yellow). \textbf{(d)} A birds-eye view of how the layer grows on GaAs and GaP substrate, the green balls represent the As and P atoms respectively.\textbf{(e)} X-ray $2\Theta/\omega$ scans of the CuMnAs \hkl(003) diffraction peak for layers grown on GaP and GaAs. The experimental data are shown as grey dots while the red line is a mosaic block model fitting for the layer grown on GaAs. This model is not applicable for the layers on GaP. The corresponding \hkl(003) reciprocal space maps are shown in the insets.\textbf{(f)} and \textbf{(g)} $Z$-contrast STEM image of CuMnAs along the \hkl[100] direction. The As positions are clearly visible as the 
brightest atoms, and intensity analysis along the rows reveal a highly ordered compound in complete agreement with the overlaid model structure in \textbf{g}. The CuMnAs unit cell, as depicted in panel \textbf{c}, is highlighted by a white dashed box in \textbf{g}.}
\label{fig:struct}
\end{figure}

\textbf{Synthesis and structure of CuMnAs epilayers.}~To stabilize the tetragonal phase of CuMnAs, layers were grown on the zincblende III-V semiconductor substrates GaAs\hkl(001) and GaP\hkl(001). In-situ reflection high energy electron diffraction (RHEED) and ex-situ x-ray diffraction (XRD) measurements, shown in Fig.~\ref{fig:struct}, revealed a structure incompatible with the chemically synthesised bulk material studied by Mundelein \etal.\cite{Mundelein:1992_a} The space group of this structure was determined using a combined experimental and computational method which enables the precise determination of the atomic positions in a thin film using conventional Cu $K\alpha$ x-ray sources only.\cite{cumnas_struct}  The space group was found to be that of the tetragonal $\mbox{Cu}_2\mbox{Sb}$ family,  P4/nmm. The lattice parameters of epitaxial CuMnAs were found to be $a$=$b$=3.820~$\mbox{\AA}$ and $c$= 6.318~$\mbox{\AA}$, for the relaxed material grown on GaAs.  

Fig.~\ref{fig:struct}c shows the unit cell of \cumnas and Fig.~\ref{fig:struct}d a birds-eye view of its registry with the substrate.  The closest match to the two substrates used, GaAs and GaP, is with the CuMnAs principal in-plane axis matching along the half diagonal of the GaAs/GaP unit cell, and this is indeed the orientation it robustly adopts for a range of temperatures and stoichiometries. In the case of the GaAs substrate the match is only preserved for the first few monolayers whereupon the layer relaxes, as evidenced by a dimming in the RHEED pattern. However it continues to grow with the same relative orientation and the RHEED pattern reappears several nanometres later. For the GaP substrate a mismatch of approximately 0.3\% is observed along the same geometry and the layer remains strained up to the thickest layers grown so far (130\,nm).

Fig.~\ref{fig:struct}e shows x-ray 2$\Theta/\omega$ linescans and reciprocal space maps (RSM) around the CuMnAs\hkl(003) diffraction peak, for \cumnas grown on GaAs\hkl(001) and GaP\hkl(001). The broadening of the RSM, as evidenced in Fig.~\ref{fig:struct}e (left inset) of the CuMnAs\hkl(003), is attributed to mosaicity introduced by the poor lattice matching of CuMnAs to the GaAs substrate. The high strain and subsequent relaxation introduce defects leading to mosaic blocks. This is supported by the fitting of a mosaic block model to the CuMnAs/GaAs which shows good agreement with a model assuming ellipsoid blocks with lateral dimensions of 50\,nm, vertical dimensions of 20\,nm and RMS misalignment of 1.5$^\circ$ (red line in Fig.~\ref{fig:struct}e).\cite{holy93,holy94}  The much sharper spot produced by the CuMnAs on GaP (right inset) indicates a more coherent layer which no longer fits a mosaic block model well.

The image shown in Fig.~\ref{fig:struct}f is a $Z$-contrast scanning transmission electron microscopy (STEM) slice along the CuMnAs \hkl[100] direction. In the enlarged region shown in Fig.~\ref{fig:struct}g the brighter As atoms are clearly visible in a triangular arrangement which matches exactly to the x-ray determined structural model along the same \hkl[100] direction, overlaid on the top right side of the image. The Cu and Mn, which are much closer in size, are not clearly distinguishable by eye, however when the intensity is profiled along the atomic planes the Cu and Mn can be distinguished and show a high degree of chemical order.

\begin{figure}[ht!]
\includegraphics[width=0.90\columnwidth]{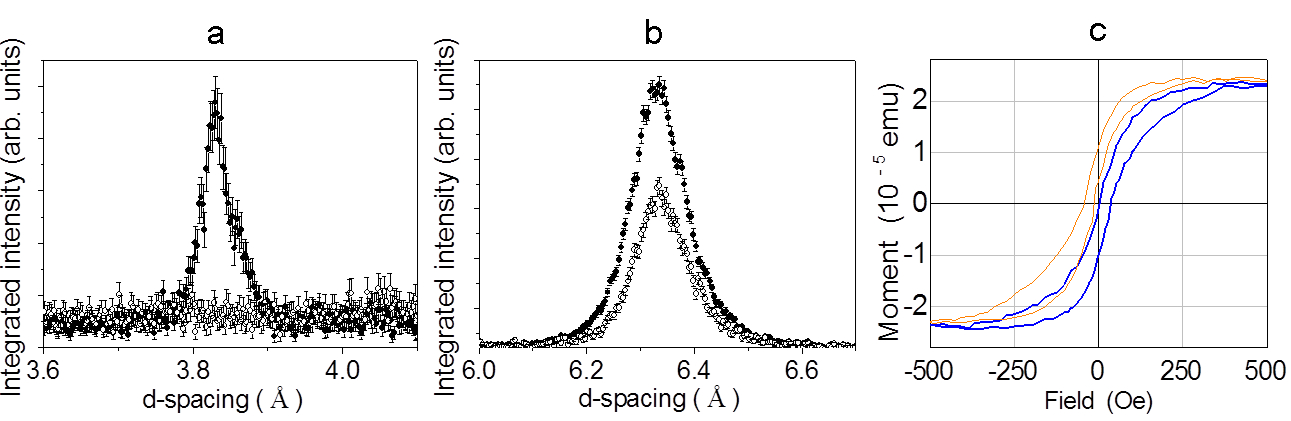}
\caption{\textbf{Magnetic characterisation.}~\textbf{(a)} Neutron diffraction data showing the structurally forbidden CuMnAs \hkl(100) diffraction peak at 300\,K (filled symbols) and 500\,K (open symbols). \textbf{(b)} Neutron diffraction from the CuMnAs \hkl(001) peak measured at the same temperatures. At 300\,K, this reflection combines both nuclear and magnetic contributions. \textbf{(c)} Hysteresis loops performed on a 3\,nm Fe film on CuMnAs at 300\,K, after cooling in a 1\,T field from 430\,K. A stable exchange bias is observed symmetrically for positive (blue) and negative (orange) field cool directions.}
\label{fig:mag}
\end{figure}

\textbf{Magnetic characterisation of CuMnAs.} Magnetic properties of the CuMnAs layers were examined using a combination of neutron diffraction and superconducting quantum interference device (SQUID) magnetometry.  Neutron diffraction measurements were taken using the WISH instrument at the ISIS neutron source. Two scattering geometries were used to access the \hkl(h00) and \hkl(00l) sets of reflections and revealed the presence of the structurally forbidden \hkl(100) magnetic reflection. The absence of this peak structurally was confirmed using in-plane x-ray diffraction (XRD), which as expected reveals the presence of the \hkl(200) and \hkl(400) reflections but not the forbidden \hkl(100). Fig.~\ref{fig:mag}a shows the \hkl(100) magnetic peak, which confirms the presence of long range AFM order with the same dimensions as the structural unit cell. 
This peak is completely absent at 500\,K, whereas the nuclear \hkl(200) peak (not shown) and the structural contribution of the \hkl(001) peak are still present (Fig.~\ref{fig:mag}b), confirming that the layer has been neither destroyed nor undergone a phase change, but has lost long range AFM order.

Analysis of the possible spin configurations, classified according to the irreducible representations associated with the $\Gamma$-point ($k=0$) of the P4/nmm space group, reveal that only models involving basis functions of the bi-dimensional $\Gamma_{5}^-$ representation are consistent with the neutron diffraction results. These models imply AFM coupling of the two symmetry related Mn-sites, with spins confined to the (\emph{ab}) plane (Fig.~\ref{fig:struct}c). Precise determination of the direction of the spin axes within this plane is complicated by the expected magnetic domains. The room temperature data normalized to a vanadium spectrum and scaled to the nuclear \hkl(200) peak  allowed us to estimate the magnetic moments of Mn to be $3.6(2)\,\mu_B$, assuming that the possible magnetic domains are equally populated.

SQUID magnetometry performed on stoichiometric CuMnAs layers revealed no ferro- or paramagnetic response, consistent with a pure AFM. In order to provide further support for the magnetic properties of the samples, CuMnAs layers were grown with 3~nm of epitaxial Fe on top and capped to prevent oxidation. By field cooling these layers through 430\,K it is possible to induce an exchange bias shift, at and above room temperature. This exchange bias is evident in the hysteresis loops shown in Fig.~\ref{fig:mag}c, and is robust and reproducible across many samples (Supplementary Fig. S1, Supplementary Fig. S2 and Supplementary Note 2). The effect remains present after training, and there is also a large change in the coercive field of the Fe layer compared to Fe on GaAs alone.\cite{wastlbauer05} Heating above 440\,K was found to damage the Fe/CuMnAs interface in a similar fashion to that observed for Fe/GaAs heterostructures,\cite{wastlbauer05,deputier97,deputier98} and after this permanent damage it is no longer 
possible to induce exchange bias (Supplementary Fig. S2a). The permanent changes to the interface were confirmed using temperature dependent x-ray reflectivity (XRR) and XRD, where roughening and density changes are observed in the interfacial region but the bulk crystal remains intact. 

\begin{figure}[!ht]
\includegraphics[width=0.45\columnwidth]{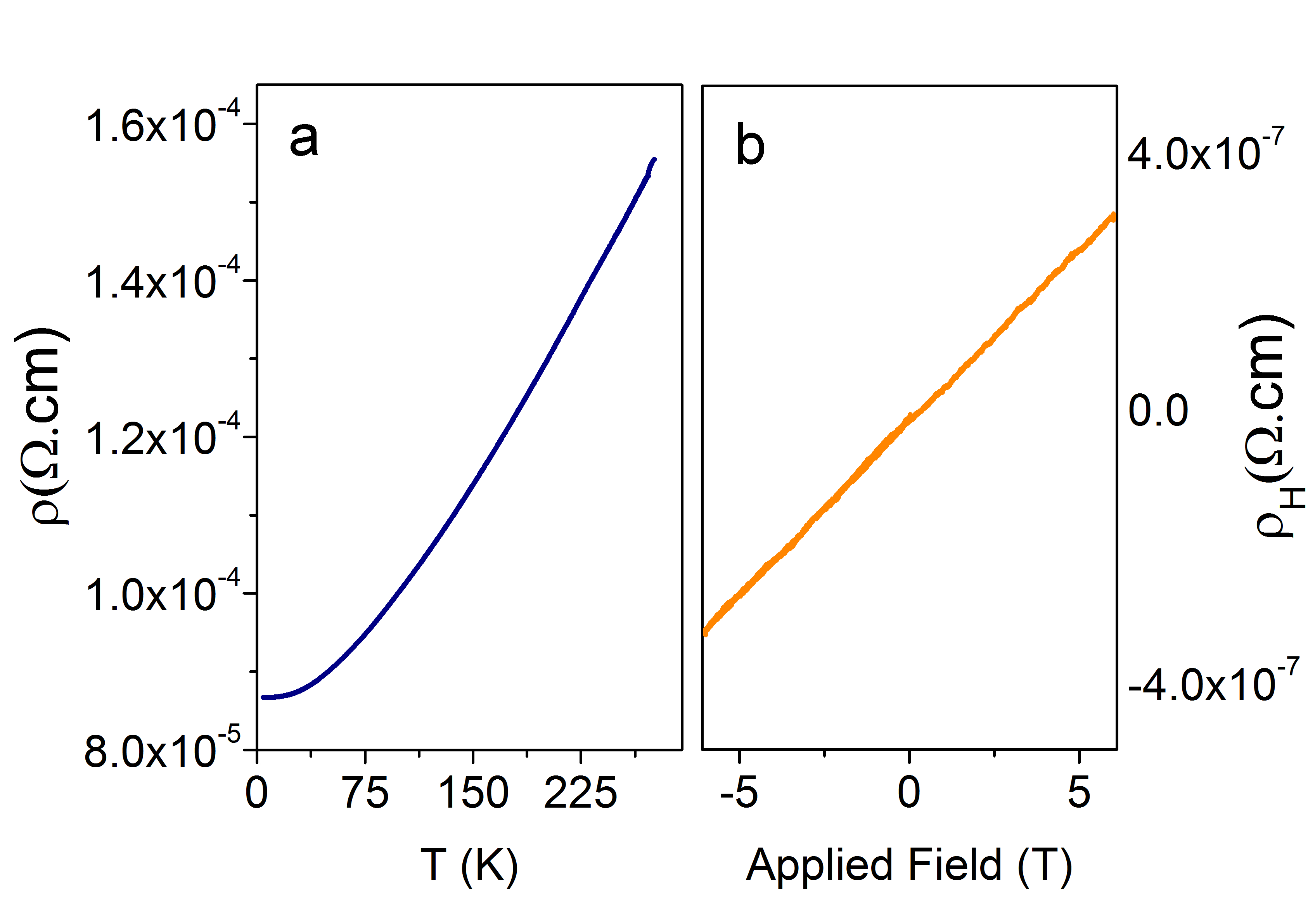}
\caption{\textbf{Transport in CuMnAs.} (a) shows the temperature dependence of resistivity and (b) the Hall resistivity as a function of applied magnetic field.}
\label{fig:tran}
\end{figure}

\textbf{Transport measurements on ${\rm\bf CuMnAs}$}. Transport measurements on stoichiometric CuMnAs samples are shown in Fig.~\ref{fig:tran}. The resistivity at 4\,K is about 90\,$\mu\Omega$.cm rising to about 160\,$\mu\Omega$.cm at 300\,K. These values are compatible with the semi-metallic-like band structure suggested by the GGA+U calculations (detailed in the next section), with strongly suppressed density of states around the Fermi energy. The measured positive Hall coefficient of $6\times10^{-8}\Omega$.cm/T interpreted in a single carrier model would give a carrier density of $1.1\times10^{22}$cm$^{-3}$. However, as discussed in Ref.~\onlinecite{Maca:2012_a} for the orthorhombic bulk CuMnAs and confirmed by our calculations for the tetragonal CuMnAs presented below, the compound's electronic structure is at the transition from a semiconductor to a semimetal. The nature of electronic states near the Fermi energy may therefore be sensitive to relatively small changes of intrinsic and extrinsic 
parameters of the CuMnAs films. In this situation, where the band structure is complex, the single carrier interpretation is inappropriate, and can greatly overestimate the number of carriers if (as the calculations indicate) the numbers of “hole-like” and “electron-like” carriers are comparable. 

\begin{figure}[ht!]
\includegraphics[width=0.90\columnwidth]{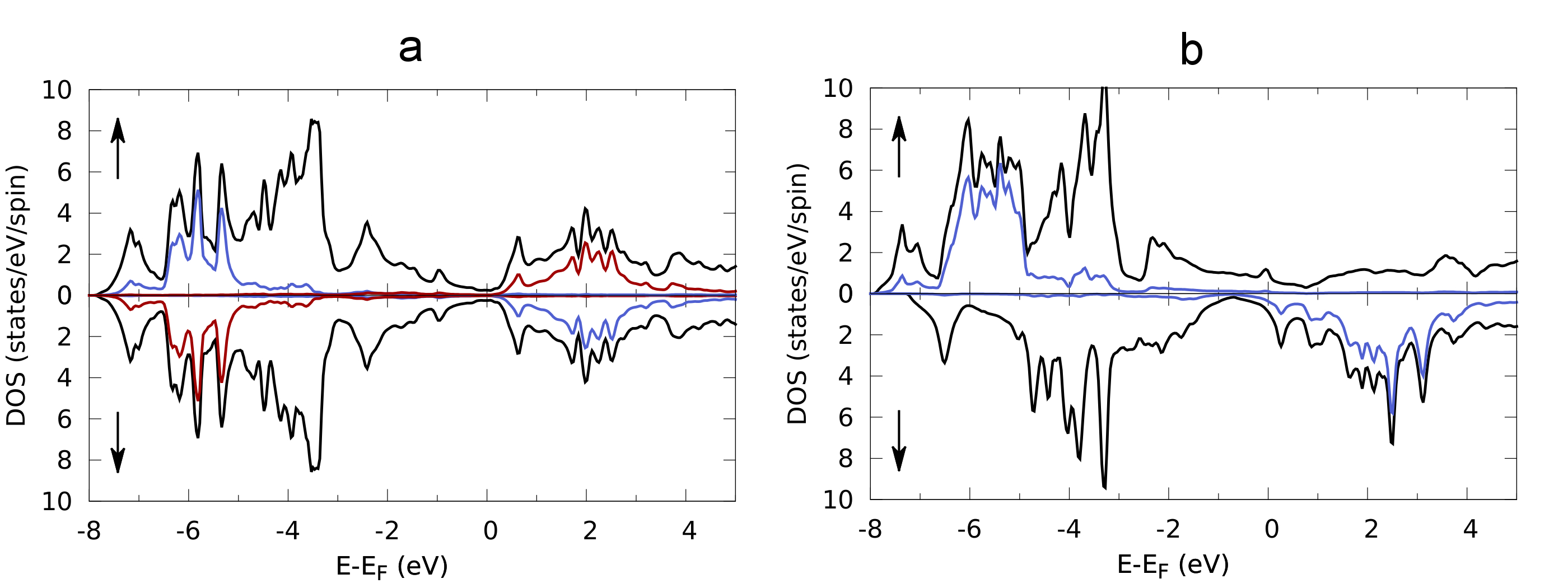}
\caption{\textbf{Density of states calculations.}~GGA+U calculated density of states of tetragonal CuMnAs for (a) antiferromagnetic and (b) ferromagnetic configurations. Upper and lower sections are for up and down spins respectively. Blue and red lines in (a) are the local density of states for Mn in each sublattice. Blue lines in (b) are the local density of states of Mn. The black lines in both panels are the total density of states.}
\label{fig:theory}
\end{figure}

\textbf{Calculations.}~From GGA+U band structure calculations, shown in Fig.~\ref{fig:theory}, we confirm that the AFM phase in CuMnAs has lower energy than the FM phase and the difference is 109 meV/f.u.. For comparison, calculations for the cubic half-Heusler CuMnSb give the difference of only 4 meV/f.u.. This suggests that the N\'eel temperature of the tetragonal CuMnAs is much higher than the $T_N=50$\,K of CuMnSb,\cite{JPSJ.29.643} as indeed is observed in our experiments. The robust AFM order is retained when the CuMnAs crystal transforms from the equilibrium bulk orthorhombic structure\cite{Mundelein:1992_a} to the tetragonal structure stabilized in our thin film epilayers. 

The GGA+U calculations show that tetragonal CuMnAs in the AFM ground state has a strongly suppressed density of states around the Fermi energy (see Fig.~\ref{fig:theory}a), indicating an electronic structure of a semimetal or a narrow bandgap semiconductor.  This is in a contrast to the FM phase which is strongly metallic (see Fig.~\ref{fig:theory}b). Our results in CuMnAs illustrate the generic trend that AFMs are more compatible with semiconductor/semimetal electronic structure than FMs. We also performed relativistic GGA+U calculations of magnetic anisotropy of the tetragonal CuMnAs in the AFM ground-state. Consistent with experiment, we find that the Mn moments, ordering ferromagnetically within the plane and antiferromagnetically between the planes, are oriented in the (\emph{ab}) plane.

Finally we note that the Ia-Mn-V compounds, NaMnAs and KMnAs, crystallize in a tetragonal structure very similar to the structure of CuMnAs with the only difference that the group Ia element occupies the Mn position in the crystal structure of CuMnAs and Mn occupies the Cu position. For this reason we compared total energies of CuMnAs when Cu and Mn switched places in the lattice. The lower energy state is observed for the configuration corresponding to the one inferred from our structural measurements of CuMnAs and shown in Fig.~\ref{fig:struct}c with the difference of 0.56 eV/f.u..

\section*{Discussion}
The results presented above demonstrate CuMnAs as a tetragonal room-temperature AFM with high crystalline quality and chemical fidelity. The exchange bias and broadening of the Fe hysteresis loop induced by CuMnAs provides not only additional evidence for the high temperature AFM ordering in the tetragonal CuMnAs epilayers, but also clearly demonstrates the potential of this new material for AFM-based spintronics functionalization. The same type of AFM/FM exchange coupling phenomena were utilized in the demonstration of the first AFM tunnelling anisotropic magnetoresistance (TAMR) devices.\cite{Park:2010_a,Marti:2012_a,Wang:2012_a}

TAMR is one of a family of spin-orbit coupling phenomena in which electronic properties and spin can influence each other. This is manifested in the realization of large magnetoresistances due to anisotropies in the density  of states and chemical potential, and in the realization of carrier-induced spin reorientation phenomena. Many of these relativistic  spintronic effects were first observed in the ferromagnetic Mn-doped zincblende GaAs. \cite{Gould:2004_a,Wunderlich:2006_a,Chernyshov:2009_a,ciccarelli12} CuMnAs shares the spin-orbit, broken space-inversion symmetry character of the electronic structure with these magnetic zincblende compounds and the tetragonal distortion of its lattice further enhances the magnetic anisotropy. The in-plane lattice parameter closely matches to GaP and Si. CuMnAs layers can be grown in a conventional III-V MBE system and therefore can be readily combined with established semiconductor heterostructure technologies. All these characteristics, complemented by the room-
temperature collinear magnetic order, make CuMnAs a favorable material candidate for exploring phenomena and functionalities within the emerging concept of spintronics based on AFMs. 

\section*{Methods}
\textbf{Growth of CuMnAs films.}~The CuMnAs films were grown in a Veeco GEN-III MBE system. Standard effusion cells were used as sources for elemental Mn and Cu, and a valved cracker cell (in the As$_4$ regime) was used as the source of As. The epitaxial growths were carried out at a substrate temperature of 300$^\circ$C, with the material fluxes set to cover a range of ratios around the stoichiometric point of Cu:Mn:As=1:1:1. For growth on GaAs\hkl(001), after opening the cell shutters the (2x4) reflection high-energy electron diffraction (RHEED) image of the clean GaAs surface became diffuse within several seconds. Typically, a new streaky (2x2) pattern emerged during the next twenty minutes (corresponding to approx. 10~nm of the epitaxial CuMnAs) which then remained stable till the end of the growth. In contrast, for growth on GaP\hkl(001) , the transition period of the diffuse RHEED image was completely missing; instead, a (1x2) pattern arose immediately after opening the shutters, transforming smoothly 
into the streaky (2x2) within the first five minutes of growth (Fig.~\ref{fig:struct}a,b).

\textbf{Structural characterisation.}~High resolution XRD measurements to determine the lattice parameters were performed on a Panalytical X'pert Materials Research Diffractometer (MRD) equipped with a standard Cu x-ray tube, a parabolic x-ray mirror, a 4x220 Ge Bartels monochromator, and a 3x220 Ge channel-cut crystal analyser with point detector. For the space group refinement a combined experimental and computational procedure, which utilised a Bruker area detector in combination with the aforementioned MRD, was used. The details of this technique are published elsewhere~\cite{cumnas_struct}.\\
High resolution aberration-corrected STEM was performed at Oak Ridge National Laboratory in a Nion UltraSTEM column, operated at 100 kV and equipped with a fifth order Nion aberration corrector. Samples were prepared by conventional thinning, grinding, dimpling and Ar ion milling.

\textbf{Calculations.}~We used the full-potential linearized-augmented-plane-wave method (WIEN2k package\cite{wien2k}) to calculate the magnetic structure of the tetragonal CuMnAs.  In all calculations we used the PBE-GGA exchange-correlation potential.\cite{wien2k} Plane wave cutoff was given by $R_{MT}*k_{max}=7.5$, where $R_{MT}$ is the smallest muffin tin radius. Brillouin zone was sampled by a mesh composed of up to 20,000 $k$-points. To describe the correlations of Mn 3$d$ electrons we employed the GGA+U approximation with $U$=4.6 eV and $J$=0.5 eV.  The lattice constants and $z$ positions of As and Mn were taken from the experiment.

\section*{Acknowledgements}
Research at UCM (M.A.R.) was supported by the ERC Starting Grant No. 23973. Research at ORNL was supported by the Materials Sciences and Engineering Division of the U.S. DOE (M.V.) and  by ORNLÕs Shared Research Equipment (ShaRE) User Facility, which is sponsored by the Office of BES, U.S. DOE. J.R. C.F. acknowledges financial support from Spanish Ministerio de Ciencia e Innovaci—n Tecnol—gica (Projects MAT2009-07967, Consolider NANOSELECT CSD2007-00041) and the Generalitat de Catalunya. X.M. acknowledges to Czech Science Foundation (Project P204/11/P339). D.Kriegner acknowledges the support by the Austrian Academy of Sciences (DOC-Program). T.J. and V.N acknowledge support from  EU ERC Advanced Grant No. 268066 and from the Ministry of Education of the Czech Republic Grants No. LM2011026, and from the Academy of Sciences of the Czech Republic Preamium Academiae. R. B. and C. R. acknowledge financial support from Fondazione Cariplo via the project EcoMag (Project No. 2010-0584). Research at the University of Nottingham was funded by EPSRC grant EP/K027808/1.

\end{document}